\documentclass[a4paper,12pt]{article}
\pdfoutput=1
\usepackage{jheppub}
\usepackage{epsf}
\usepackage{epsfig}
\usepackage{subfigure}
\usepackage{mathtools}
\usepackage{hhline}
\usepackage{float}
\usepackage{multirow}
\usepackage{nicefrac}
\usepackage{epstopdf}
\usepackage{slashed}
\usepackage{xcolor}
\usepackage{url}
\usepackage{umoline}
\usepackage{graphicx,subfigure}
\usepackage[utf8x]{inputenc}
\usepackage[normalem]{ulem}
\newcommand{\be}{\begin{equation}}
\newcommand{\ee}{\end{equation}}
\newcommand{\bes}{\begin{equation*}}
\newcommand{\ees}{\end{equation*}}
\newcommand{\bea}{\begin{eqnarray}}
\newcommand{\eea}{\end{eqnarray}}
\newcommand{\beas}{\begin{eqnarray*}}
\newcommand{\eeas}{\end{eqnarray*}}

\newcommand{\damu}{$\delta a_\mu~$\!}
\newcommand{\dae}{$\delta a_e~$\!}
\newcommand{\tb}{\tan\beta~}

\title{Explaining $g-2$ anomalies in two Higgs doublet model with vector-like leptons}
\author{Eung Jin Chun,}
\author{Tanmoy Mondal}
\affiliation{Korea Institute for Advanced Study, Seoul 02455, Korea}
\emailAdd{ejchun@kias.re.kr}
\emailAdd{tanmoy@kias.re.kr}

\abstract{
We consider the two Higgs doublet model (2HDM)  along with a generation of vector-like lepton doublet and singlet
to explain the observed discrepancies in the electron and muon anomalous magnetic moments. The type-X (lepton-specific)
2HDM can allow  a light pseudo-scalar which is known to explain the muon anomalous magnetic moment at two-loop.  Such a
light particle  induces a sizable negative contribution to the electron anomalous magnetic moment at one-loop in the
presence of vector-like leptons evading all the experimental constraints.
}

\preprint{KIAS - P20050}
\date{\today}

\begin{document}
\maketitle

\section{Introduction}
The precise determination of electron and muon anomalous magnetic moment, both theoretical and experimental, is an
important test for the Standard Model(SM). Present experimental value~\cite{Bennett:2006fi} of the muon anomalous
magnetic moment ($a_\mu=(g-2)_\mu$) indicates a $3.7~\sigma$ deviation from the SM
prediction~\cite{Blum:2018mom,Keshavarzi:2018mgv,Davier:2019can,Aoyama:2020ynm}\footnote{See also~\cite{Campanario:2019mjh}}:
\be
\delta a_\mu = a_\mu^{\rm{EXP}}-a_{\mu}^{\rm{SM}} = \left(2.706\;\pm\;0.726\right)\;\times 10^{-9}.
\ee
On the other hand, the measurement of fine structure constant using the Cs atom gives a precise estimation of the
electron anomalous magnetic moment ($a_e=(g-2)_e$) in the SM~\cite{Parker:2018vye} which is $2.4~\sigma$ below the
experimentally measured value~\cite{Hanneke:2008tm}:
\be
\delta a_e = a_e^{\rm{EXP}}-a_e^{\rm{SM}} = -\;\left(8.8\;\pm\;3.6\right)\;\times 10^{-13}.
\ee
Intriguingly, the deviations are in opposite direction which makes it hard to explain in a unified way. Also, the
absolute value of the deviations are not proportional to the square of the respective lepton mass and it is unlikely
that both the anomalies share a common origin.  Various suggestions have been made for the simultaneous explanation
of the anomalies~\cite{Giudice:2012ms,Davoudiasl:2018fbb,Crivellin:2018qmi,Liu:2018xkx,Han:2018znu,Endo:2019bcj,Abdullah:2019ofw,Bauer:2019gfk,Badziak:2019gaf,Hiller:2019mou,Cornella:2019uxs,CarcamoHernandez:2020pxw,Haba:2020gkr,Bigaran:2020jil,Jana:2020pxx,Calibbi:2020emz,Chen:2020jvl,Yang:2020bmh,Hati:2020fzp,Dutta:2020scq,Botella:2020xzf,Chen:2020tfr,Dorsner:2020aaz,Arbelaez:2020rbq,Jana:2020joi}.

The two Higgs doublet model (2HDM) is one of the simplest extension of the SM scalar sector, where two scalar doublets
are involved in the electroweak symmetry breaking.  An extra Higgs doublet appears in many theories beyond the
SM: supersymmetry~\cite{Haber:1984rc}, explaining the baryon asymmetry of the
Universe~\cite{Turok:1990zg,Trodden:1998ym}, or resolving the strong CP problem~\cite{Kim:1986ax}.
The presence of a relatively light neutral scalar in 2HDM does not violate the custodial symmetry~\cite{Gerard:2007kn}
and thus can be consistent with the electroweak precision test~\cite{Broggio:2014mna}.
It is well known that the type-X (lepton-specific) 2HDM with a light pseudo-scalar ($A$)
can explain the muon anomaly for large
$\tan\beta$~\cite{Cao:2009as,Broggio:2014mna,Wang:2014sda,Ilisie:2015tra,Abe:2015oca,Han:2015yys,Chun:2016hzs,Cherchiglia:2016eui,Cherchiglia:2017uwv,Wang:2018hnw}.
On the other hand, the contribution to the electron $(g-2)$ from the type-X 2HDM scales as $(m_e/m_\mu)^2$ and remains
small. Moreover, the contribution has the same sign as the muon anomaly. Consequently, pure 2HDM  type-X can not
explain both anomalies.

Hence to explain the electron anomaly a larger contribution to \dae\!  is needed compared to the muon case.
For this, we consider the inclusion of new vector-like fermions which couple dominantly to the electron.
Vector like fermions have already been studied extensively and are motivated in several BSM models, including
extra dimension and grand unified theories~\cite{Hewett:1988xc,Thomas:1998wy}. In particular, additional new
vector-like lepton (VLL) can help to explain the anomalous magnetic moment since the helicity flip which is
required by the dipole transition occurs through a VLL mass insertion which is larger than the electroweak (EW) scale.
The SM model with a singlet and a doublet VLL which couple exclusively to the muon has been suggested to explain
\damu~\cite{Kannike:2011ng,Dermisek:2013gta}.
Moreover, the VLL mass, if close to the EW scale, modifies significantly the muon Yukawa coupling
which is strongly constrained by the recent measurements of $h\to\mu\mu$ by ATLAS~\cite{Aad:2020xfq} and
CMS~\cite{CMS:2020eni}. For recent studies to explain muon $(g-2)$ in 2HDM with vector-like lepton see~\cite{Barman:2018jhz,Frank:2020smf}.

In this paper, we explore this idea in the context of  the two Higgs doublet model.
In particular we consider the type-X  2HDM where one doublet ($\Phi_2$) is responsible for the masses of the quarks
while the other ($\Phi_1$) generates masses of the leptons. The new vector-like leptons are supposed to couple
dominantly to the electron through Yukawa couplings with $\Phi_1$. Since the new vector-like leptons couple to the doublet
which acquires a small vacuum expectation value ($vev$), the corrections to the SM
couplings will be small. We find that the model remains mostly unconstrained from both the precision measurements
at the $Z$ pole and the oblique corrections. In this scenario, \dae is generated at one-loop owing to the mixing
of the new leptons with the electron. The dominant contribution  comes from the pseudo-scalar ($A$) mediated diagram.
Apart from explaining the \dae, the VLL  contributes also to \damu at two-loop which is, however, relatively small
as it is suppressed by the ratio of the small chiral mass to the vector-like mass of the new leptons.
Nevertheless, this additional contribution can help explain \damu in a larger parameter space compared to
the pure type-X 2HDM scenario.

Our model predicts a  peculiar collider signature.
A pair of the vector-like leptons produced at the LHC will leave an electron-positron pair and extra Higgs bosons,
especially, a light pseudo-scalar $A$ which will be boosted and decay to $\tau^+\tau^-$.
Thus the over-lapping di-$\tau$ accompanied by $e^+ e^-$ can be searched for to probe our scenario.
The novel di-$\tau$ tagger used by the ATLAS collaboration for di-Higgs searches will be useful also for the
boosted $A$ search.

The paper is organized as follows: In Sec.~\ref{sec:model} we describe the model in detail.
Then in Sec.~\ref{sec:el-mu-g-2} we move on to the contribution to electron and muon anomalous magnetic moment
in this model. The constraints coming from $Z$ pole precision physics and oblique parameters are discussed in
Sec.~\ref{sec:precision-obs}. The results are described in Sec.~\ref{sec:result} and possible collider bounds and
signals are discussed in Sec.~\ref{sec:collider}. Finally we conclude in Sec.~\ref{sec:conclusion}.

\section{The Model}\label{sec:model}
\begin{table}[t]
\begin{center}
\begin{tabular}{|c||c|c|c|c|c|c|}
\hline
Fields       & $\ell_L$ & $e_R$ & $L_L,L_R$ & $E_R, E_L$ & $\Phi_1$ & $\Phi_2$ \\ \hline
$Z_2$ Charge & +        & $-$     & +         & $-$          & $-$        & +        \\ \hline
\end{tabular}
\end{center}
\caption{Table contains the  $Z_2$ charges of leptons and scalars present in the model. }
\label{tab:z2_charge}
\end{table}
We consider a model with 2HDM (of type-X) along with vector-like leptons. That is, the SM is extended to have
an additional scalar doublet, a pair of lepton doublets $L_{L,R}$ and a pair of charged lepton singlets $E_{L,R}$.
The two doublets $\Phi_1$ and $\Phi_2$ have same hypercharge ($+\frac12$). To avoid flavor changing neutral
current processes we have considered additional $\mathbb{Z}_2$ symmetry under which $\Phi_1$ is odd and $\Phi_2$ is even.
The SM singlet leptons are odd under the $\mathbb{Z}_2$ symmetry which ensures that the leptons couples solely to $\Phi_1$.
In the same spirit, we have considered that the singlets $E_{L,R}$ are odd under the $\mathbb{Z}_2$ symmetry.
In Tab.~\ref{tab:z2_charge} we show leptons and scalars under the  $\mathbb{Z}_2$ symmetry.
Also, we have considered the mixing of VLL with electrons only and  mixing of new leptons  with more
than one SM family simultaneously is strongly constrained by various lepton flavor violating processes.

Relevant part of the Lagrangian containing Yukawa and mass terms(suppressing the generation index):
\bea\label{eq:lagrangian}
-\mathcal{L} &\supset& \;  y_{e} \; \bar \ell_{L} e_{R} \Phi_1 + \lambda_L\; \bar L_{L}  e_{R}  \Phi_1  + \lambda_E \; \bar \ell_{L}  E_{R} \Phi_1\nonumber \\
&& +  \lambda \; \bar L_{L}  E_{R}  \Phi_1 + \bar \lambda\; \Phi_1^\dagger \bar E_{L} L_{R} + M_L\; \bar L_L L_R + M_E\; \bar E_L E_R + {\rm h.c.}~.
\eea
Let us note that  two additional mass terms $\mu_L \bar\ell_L L_R$ and $\mu_E \bar E_L e_R$ allowed in the Lagrangian
can be rotated away, so that only the
vector-like mass terms ($M_L$ and $M_E$) and Yukawa couplings remain as free parameters.

The lepton and scalar doublets can be written as,
\be  \ell_L  = \left( \begin{array}{c} \nu_e \\ e^-_L \end{array} \right),~ L_{L,R}  = \left( \begin{array}{c} L_{L,R}^0 \\
 L_{L,R}^-\end{array}\right),
 ~\Phi_1  = \left( \begin{array}{c} \Phi_1^+ \\ \Phi_1^0 \end{array} \right),
 ~\Phi_2  = \left( \begin{array}{c} \Phi_2^+ \\ \Phi_2^0 \end{array} \right).
 \ee
As usual in 2HDM, we have,
\bea\label{eq:phi10_phi20}
\Phi_1^0 &=& \dfrac{1}{\sqrt2}\Big(v_1(=v c_\beta)+c_\alpha\, H -s_\alpha\, h + i\,c_\beta\,G^0 -i\,s_\beta\,A\Big)\nonumber\\
\Phi_2^0 &=& \dfrac{1}{\sqrt2}\Big(v_2(=v s_\beta)+s_\alpha\, H +c_\alpha\, h + i\,s_\beta\,G^0 +i\,c_\beta\,A\Big),
\eea
where $h$ is the SM higgs boson and $H(A)$ is additional scalar(pseudo-scalar).
The charged gauge eigenstates $\Phi_1^+$
and $\Phi_2^+$ will give rise to one charged Higgs $H^+$ and a charged Goldstone boson.
 Details about the scalar sector in 2HDM can be found in~\cite{Branco:2011iw}.

After spontaneous symmetry breaking the  mass matrix for charged leptons is given by:
\be
\mathcal{L}_{mass} = ( \bar \ell_{Li}, \bar L^-_L, \bar E_L ) \;
\mathcal{M}_E \;  \begin{pmatrix}
 \ell_{Rj} \\
 L^-_R\\
 E_R
\end{pmatrix} + h.c..
\ee
 Here $i,j$ denotes the light lepton generation index ($i,j=1,2,3$) and  $\mathcal{M}_E$ is the  $5\times 5$ mass matrix given by
\be
\mathcal{M}_E =
\begin{pmatrix}
  \frac{1}{\sqrt2}\;y_{e,ij}\; v_1 & 0 &  \frac{1}{\sqrt2}\;\lambda_{E_i}\;v_1 \\
 \frac{1}{\sqrt2}\; \lambda_{L_j}\;v_1 & M_L &  \frac{1}{\sqrt2}\;\lambda\;v_1\\
 0 & \frac{1}{\sqrt2}\;\bar \lambda \;v_1 & M_E
\end{pmatrix},
\label{eq:mass_matrix}
 \ee
where the upper left block is the $3\times 3$ matrix of the SM leptons.
The mass matrix  can be diagonalized by a bi-unitary transformation,
\be\label{eq:mass-diag}
\widetilde U^\dagger_L\; \mathcal{M}_E \;\widetilde U_R = {diag}(m_e, m_\mu, m_\tau, m_1, m_2),
\ee
where $m_1$ and $m_2$ denotes mass of the two new mass eigenstates $E_1$ and $E_2$ respectively.
 Since we are interested in the mixing of electron with the vector-like leptons, we have 
assumed that the $\lambda_{E_2},\lambda_{E_3},\lambda_{L_2},\lambda_{L_3}=0$. From now on we will 
ignore the muon and tau in the charged lepton mass matrix $\mathcal{M}_E$ and denote 
$\lambda_{E_1(L_1)}=\lambda_{E(L)}$. With these simplifications we can now rewrite the 
Eq.~\ref{eq:mass-diag} as,
\begin{eqnarray}
U^\dagger_L
\begin{pmatrix}
  \frac{1}{\sqrt2}\;y_{e}\; v_1 & 0 &  \frac{1}{\sqrt2}\;\lambda_{E}\;v_1 \\
 \frac{1}{\sqrt2}\; \lambda_{L}\;v_1 & M_L &  \frac{1}{\sqrt2}\;\lambda\;v_1 \\
 0 & \frac{1}{\sqrt2}\;\bar \lambda \;v_1 & M_E
\end{pmatrix}
U_R
 =
 \begin{pmatrix}
m_e  & 0 &   0\\
 0 & m_{1} &  0\\
 0 & 0 & m_{2} \\
\end{pmatrix}.
\label{eq:mass}
\end{eqnarray}

 In the above equation the diagonalization matrices $U_{L,R}$ obtained from $\widetilde U_{L/R}$ by removing the 
muon and tau entries.

 In the limit
 \begin{equation}
\frac{\lambda_L v_1}{\sqrt2},~~\frac{\lambda_E v_1}{\sqrt2},~~\frac{\bar \lambda v_1}{\sqrt2},~~\frac{\lambda v_1}{\sqrt2} ~ \ll~ M_E, M_L
\label{eq:integrateout}
 \end{equation}
approximate analytic formulas for diagonalization matrices can be obtained~\cite{Grimus:2000vj,Dermisek:2013gta},
\begin{equation}
U_{L} =  \renewcommand\arraystretch{1.5}  \left(\begin{array}{ccc}
1 - \frac{v_1^2}{2} \frac{\lambda_E^2}{2M_E^2}
& -\frac{v_1^2}{2} \left( \frac{\lambda_E}{M_L}\frac{ \bar{\lambda} M_E + \lambda M_L}{M_E^2 - M_L^2} - \frac{y_e \lambda_L}{M_L^2}\right)
& ~~\frac{v_1}{\sqrt2}\frac{\lambda_E}{M_E} \\
\frac{ v_1^2 (\bar{\lambda} \lambda_E M_L  -  y_e \lambda_L M_E  )}{2 \; M_L^2 M_E}
& \cos\theta_L
&  -\sin\theta_L \\
- \frac{v_1}{\sqrt2}\frac{\lambda_E}{M_E}
& \sin\theta_L
& \cos\theta_L   \end{array}   \right),
\label{eq:UL}
 \end{equation}
 and
\begin{equation}
U_R =  \renewcommand\arraystretch{1.9} \left(\begin{array}{ccc}
1 - \frac{v_1^2}{2} \frac{\lambda_L^2}{2M_L^2}
  & \frac{v_1}{\sqrt2} \frac{\lambda_L}{M_L}
  &~~ \frac{v_1^2}{2} \left(\frac{\lambda_L}{M_E} \frac{ \bar{\lambda} M_L + \lambda M_E}{ M_E^2 - M_L^2} + \frac{y_e \lambda_E}{M_E^2}  \right)   \\
-\frac{v_1}{\sqrt2}\frac{\lambda_L}{M_L}
  & \cos\theta_R
  & -\sin\theta_R \\
\frac{v_1^2}{2}\frac{ (\lambda_L  \bar{\lambda} M_E  -  y_e \lambda_E M_L)}{M_L M_E^2}
  &   \sin\theta_R
  &  \cos\theta_R
\end{array} \right),
\label{eq:UR}
\end{equation}
where the $2\times2$ matrix $R(\theta_{L/R})$ diagonalizes the 2-3 block of
$\mathcal{M}_E\mathcal{M}_E^\dagger$ and $\mathcal{M}_E^\dagger\mathcal{M}_E$ respectively.

\subsection{Couplings of the scalars and fermions}\label{sec:All-coup}
From Eqs.~\ref{eq:lagrangian}
and~\ref{eq:phi10_phi20} we can write the coupling of the  electron and the new vector-like fermions with the pseudo-scalar as
\begin{eqnarray}\label{eq:a-coup}
\mathcal{L}_{A\chi_a\chi_b}&=&-i\;\frac{1}{\sqrt2}\;s_\beta \; A \; (e_L,\;L^-_L,\;E_L) \;\begin{pmatrix}
 y_e & 0 &  \lambda_E \\
\lambda_L  &0 &  \lambda \\
 0 & \bar \lambda  & 0
\end{pmatrix} \;
\begin{pmatrix}
 e_R\\L_R^-\\E_R
\end{pmatrix} + h.c.\\
 \label{eq:xi-ab}
&=& -i \tan\beta\, {\xi_{ab} \over v}  \; A \; \bar\chi_{La}  \chi_{Rb} + h.c.,
\end{eqnarray}
where $\chi_{L/R} = (e,E_1,E_2)_{L/R}$  are the mass eigenstates obtained from diagonalization by $U_{L/R}$
 and  the mass matrix $\xi_{ab}$ is given by
\be
\xi_{ab}=\left[ {diag}(m_e,\,m_1,\,m_2) - U_L^\dagger\;{diag}(0,M_L,M_E)\;U_R\right]_{ab}.
\ee
Similarly, one obtains the Yukawa couplings of the scalars $H$ and $h$ as
\be\label{eq:Hh-coup}
\mathcal{L} \supset ~~ \dfrac{c_\alpha}{c_\beta} \; \dfrac{1}{v} \;\xi_{ab}\;H\;\bar\chi_{La} \chi_{Rb} ~~ - ~~
\dfrac{s_\alpha}{c_\beta} \; \dfrac{1}{v} \;\xi_{ab}\; h \;\bar\chi_{La} \chi_{Rb}+ h.c.
\ee
From the expression of $\xi_{ab}$ it is evident that if there are no new fermion fields then $\xi_{ab}$ 
reduces to the mass of the lepton ($\xi_{11}=m_e$) and the standard 2HDM Yukawa couplings are recovered:
\be\label{eq:All-type-X}
\mathcal{L} \supset -i y_\ell^A \;\left(\dfrac{m_\ell}{v}\right)\;A \;\bar \ell \gamma_5\ell + 
  y_\ell^H \; \left(\dfrac{m_\ell}{v}\right)\;H \;\bar \ell \ell + 
  y_\ell^h \left(\dfrac{m_\ell}{v}\right)\;  \;h \;\bar \ell \ell.
\ee
with the prefactors $y^{A,H,h}_l$ specific in the type-X model:
\be
y_\ell^A = \tb, \hspace{1cm}y_\ell^H =  \dfrac{c_\alpha}{c_\beta} \hspace{1cm}\textrm{and}\hspace{1cm} y_\ell^h =  \dfrac{-s_\alpha}{c_\beta}.
\ee
To denote coupling of $A$ and $H$ with electron and the vector-like leptons, 
    we will use convention of Eq.~\ref{eq:xi-ab} and \ref{eq:Hh-coup}. On the other hand, for muon and tau we will use the notation of 
    Eq.~\ref{eq:All-type-X}. This is to emphasize the fact that only the electron coupling is affected by VLLs.

\subsection{Couplings to Gauge bosons}

Apart from the pseudo-scalar mediated diagram, there will be an additional diagram for \dae mediated by the SM
gauge bosons. Here we give explicit expressions relevant for the computation of \dae.
The couplings of $Z$ boson with electron and the heavy fermions is given by,
\bea\label{eq:z-coup}
g_L^{Z\,e\,E_1} &=& \dfrac{g}{2\cos\theta_W} (U_L^\dagger)_{13} (U_L)_{32} {\textrm{~,~~~~~}}
g_L^{Z\,e\,E_2} = \dfrac{g}{2\cos\theta_W} (U_L^\dagger)_{13} (U_L)_{33}\nonumber\\
g_R^{Z\,e\,E_1} &=& -\dfrac{g}{2\cos\theta_W} (U_R^\dagger)_{12} (U_R)_{22} {\textrm{~,~~~~~}}
g_R^{Z\,e\,E_2} = -\dfrac{g}{2\cos\theta_W} (U_R^\dagger)_{12} (U_R)_{23}
\eea
Similarly, the coupling of the $W$ boson with electron and the neutral heavy lepton($N$) is given by
\be\label{eq:w-coup}
g_L^{WN} = \dfrac{g}{\sqrt2}(U_L)_{21} {\textrm{~,~~~~~}} g_R^{WN} = \dfrac{g}{\sqrt2}(U_R)_{21}
\ee

Note that all the couplings are proportional to the $v_1$ and hence the gauge boson couplings   are always
much smaller than the $A-VLL-\textrm{electron}$ couplings. This is important to satisfy the precision constraints.

\section{Electron and muon anomalous magnetic moment}\label{sec:el-mu-g-2}

\subsection{Electron $(g-2)$}\label{subsec:el-g-2}

\begin{figure}
\centering
\includegraphics[width=7cm]{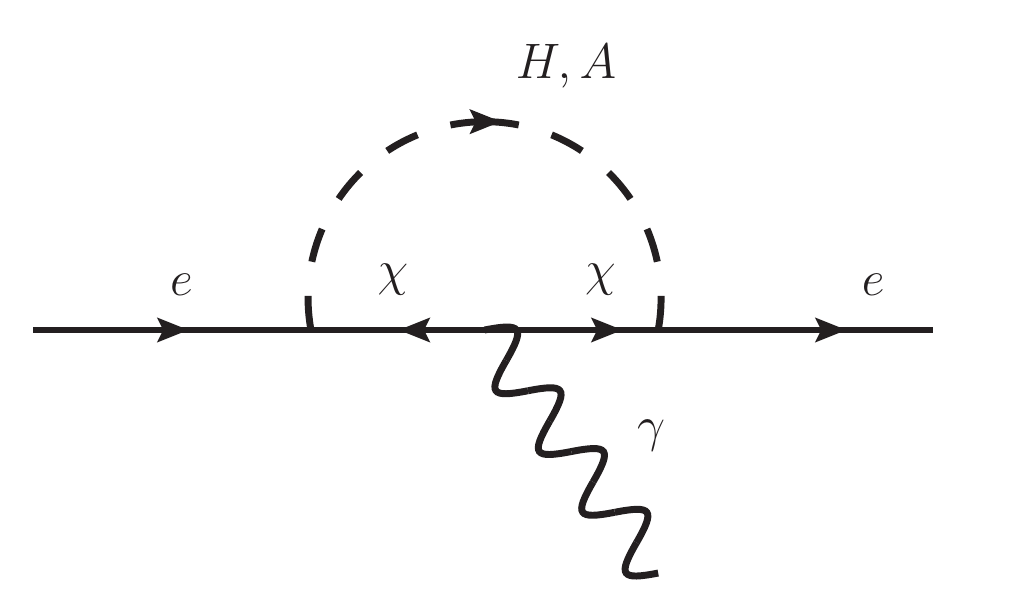}
\includegraphics[width=7cm]{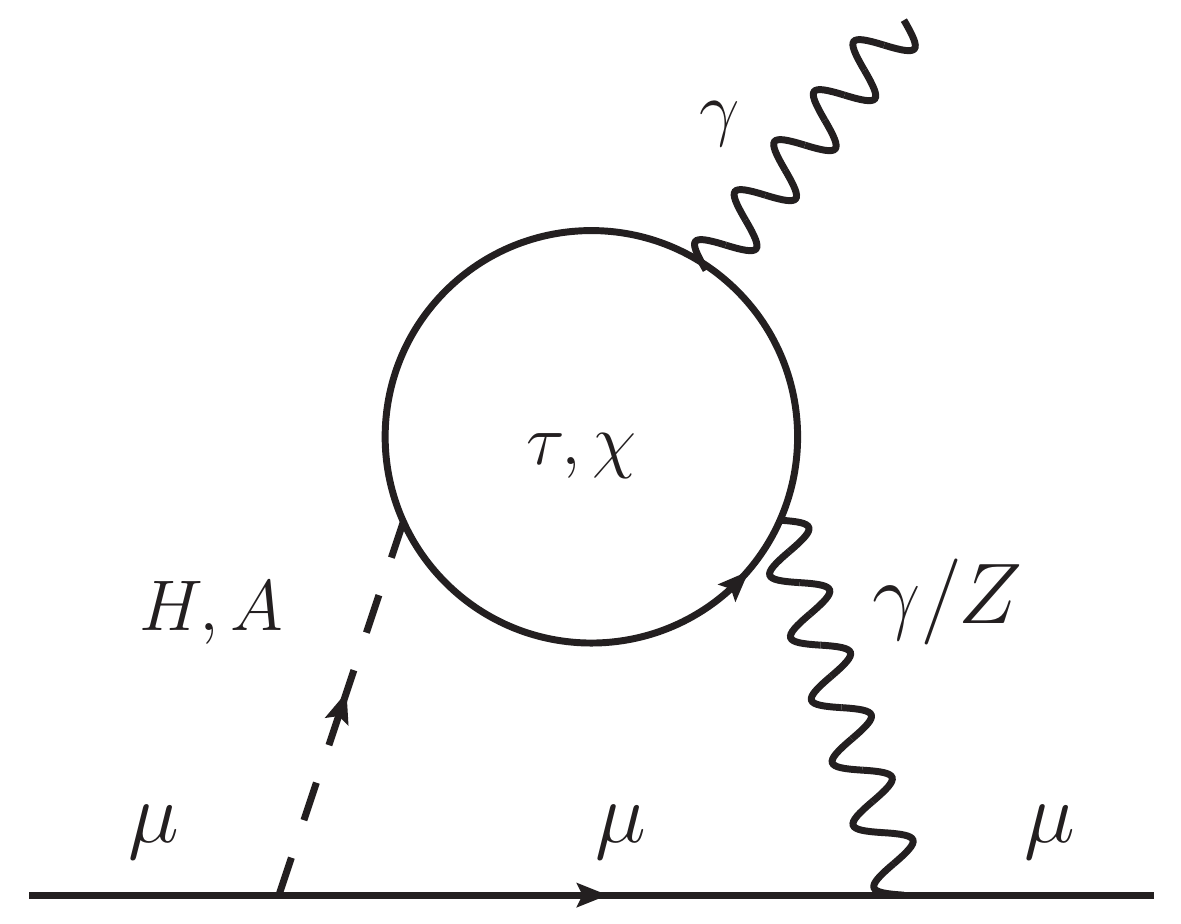}
\caption{Left figure shows electron $(g-2)$ dominant contribution from the chirality flip of the heavy leptons
and the right figure is a representative two-loop Barr-Zee diagram which explain the muon $(g-2)$.
Here $\chi$ denotes the vector-like leptons.}
\label{fig:electron_g-2}
\end{figure}

The dominant contribution to the anomalous magnetic moment will come from diagrams with a chiral flip of the VLL
in the loop as shown in Fig.~\ref{fig:electron_g-2}. The contribution from the pseudo-scalar $A$ in the loop is given
by~\cite{Lindner:2016bgg},
\be\label{eq:diag-A-med}
\delta a_e^A = -\dfrac{m_e^2}{8~\pi^2~m_A^2}\left(\;\dfrac{\tan\beta}{v}\right)^2
    \sum_{i=1,2}\;\; \xi_{1i}\;\xi_{i1}\;I_-\left(\dfrac{m_e^2}{m_A^2},\;\dfrac{m_i^2}{m_A^2}\right) ,
\ee
 and  the  heavy neutral Higgs $H$ mediated diagram gives,
\be\label{eq:diag-H-med}
 \delta a_e^H = -\dfrac{m_e^2}{8~\pi^2~m_H^2}\left(\dfrac{c_\alpha}{c_\beta} \; \dfrac{1}{v}\right)^2
    \sum_{i=1,2}\; \xi_{1i}\;\xi_{i1}\;I_+\left(\dfrac{m_e^2}{m_H^2},\;\dfrac{m_i^2}{m_H^2}\right)   .
\ee
The loop functions are,
\be
I_{\pm}(a,b)=\int_0^1 dx\;\frac{x^2(1-x\pm\epsilon)}{(1-x)(1-x\;a)+x\,b}\textrm{~~where~~}\epsilon=\frac{m_i}{m_e}.
\ee
There are also diagrams with the charged Higgs boson and heavy neutral lepton in the loop.
However, their contribution becomes negligible due to lack of a chiral enhancement.

The contribution from the $Z$ diagram is given by
\begin{equation}
\delta a_e^Z = -\frac{m_e}{8 \pi^2 m_Z^2} \sum_{i=1,2}  \left[  ( g^{Z\, e\,E_i}_L g^{Z\, e \,E_i}_R ) \, m_{i}~  G_Z(x_{Z, i})  \right] , \end{equation}
where $x_{Z,i} = (m_{i}/M_Z)^2$, the couplings are expressed in Eq.~\ref{eq:z-coup} and the loop function is as follows:
\begin{eqnarray}
G_Z(x) &=&  \frac{x^3  + 3 x  - 6 x \ln(x) - 4}   { 2 (1 - x)^3}.
\end{eqnarray}
The $W$ mediated diagram yields,
\bea
\delta a_e^W = -\frac{m_e}{16 \pi^2 m_W^2 }   \left[( g_L^{W N} g_R^{W N} ) \, M_L G_W(x_{W})  \right],
\eea
where $x_{W} = (M_L/M_W)^2$. The couplings are written in Eq.~\ref{eq:w-coup}
and the loop function is:
\bea
G_\text{W}(x) &=&  -\frac {x^3 - 12 x^2 + 15 x  + 6 x^2 \text{ln} (x)  -4   }  { (1 - x)^3 }.
\eea
Contribution from gauge boson  diagrams is much smaller than the pseudo-scalar one since
this couplings are $\tb$ suppressed compared to $A$.

Please note that we have not included the diagrams which contain only the SM particles as they are
in principle included in the EW contribution from the Standard Model. These pure SM diagrams do not
contribute substantially due to lack of any chiral enhancement.

\subsection{Muon $(g-2)$}\label{subsec:mu-g-2}

The vector-like leptons do not couple to muons, and there are no new VLL loops for muon $(g-2)$ at one loop.
The one-loop diagrams are subdominant compared to the two-loop Barr-Zee (BZ) diagrams with heavy fermions($\tau$ or VLL)
in the loop as shown in right panel of Fig.~\ref{fig:electron_g-2}.
The new vector-like leptons will give an additional contribution to the muon ($g-2$).
The dominant contribution reads as
\bea\label{eq:damu-A}
\delta a_\mu^A = \frac{\alpha_{em}}{4\pi^3} \frac{m_\mu^2}{v^2}& \Big[&
\sum_{i=1,2} Q_{E_i}^2 ~y_\mu^A \frac{v}{m_i}\; g_{A\,E_i\,E_i} \;\mathcal{F}\left(\frac{m_{i}^2}{m_A^2}\right)
+ Q_\tau^2\; y_\mu^A \; y_\tau^A  \;\mathcal{F}\left(\frac{m_\tau^2}{m_A^2}\right)\Big].
\eea

Relevant contribution coming from the heavy neutral higgs is given by
\bea\label{eq:damu-H}
\delta a_\mu^H = \frac{\alpha_{em}}{4\pi^3} \frac{m_\mu^2}{v^2}& \Big[&
\sum_{i=1,2} Q_{E_i}^2 ~y_\mu^H \frac{v}{m_i}\; g_{H\,E_i\,E_i} \;\mathcal{G}\left(\frac{m_{i}^2}{m_H^2}\right)
+ Q_\tau^2\; y_\mu^H \; y_\tau^H  \;\mathcal{G}\left(\frac{m_\tau^2}{m_H^2}\right)\Big].
\eea

The loop functions are,
\bea
\mathcal{F}(x)&=&\frac{x}{2}\int_0^1\,dy~\frac{1}{y(1-y)-x}\;ln\left(\frac{y(1-y)}{x}\right),\\
\mathcal{G}(x)&=&\frac{x}{2}\int_0^1\,dy~\frac{2y(1-y)-1}{y(1-y)-x}\;ln\left(\frac{y(1-y)}{x}\right).
\eea
The factors $g_{A\,E_i\,E_i}$ and $g_{H\,E_i\,E_i}$ can be calculated from Eq.~\ref{eq:xi-ab} and \ref{eq:Hh-coup}:
\be
g_{A\,E_i\,E_i} = \tan\beta\; \dfrac{1}{v} \;\xi_{ii} \textrm{~~~~~and~~~~~} 
g_{H\,E_i\,E_i} = \dfrac{c_\alpha}{c_\beta}\; \dfrac{1}{v}\;\xi_{ii}.
\ee
The factors $y^{A/H}_{\mu/\tau}$ are  defined in Eq.~\ref{eq:All-type-X}.
Contribution coming from the tau-loop is shown in the second term of 
Eq.~\ref{eq:damu-A} and \ref{eq:damu-H} and is same as in type-X 
2HDM~\cite{Cao:2009as,Broggio:2014mna,Wang:2014sda,Ilisie:2015tra,Abe:2015oca,Han:2015yys,Chun:2016hzs,Cherchiglia:2016eui,Cherchiglia:2017uwv,Wang:2018hnw}.
As discussed previously, we have used 2HDM conventions for muon and tau loop because only electron coupling is affected by VLLs.

The Yukawa modifiers $y_\mu^H$  and $y_\tau^H$ has the same value $\dfrac{c_\alpha}{c_\beta}$
which goes as $\tb$ in the limit $\sin(\beta-\alpha)\simeq 1$ as indicated by the Higgs measurements at the
LHC~\cite{Haller:2018nnx}. When the heavy Higgs is lighter than the vector like particle in the loop then the
contribution is not suppressed by the heavy Higgs mass and nearly comparable to the light pseudo-scalar contribution.
Moreover, the contribution from $H$ mediated diagram is negative and partially cancels the $A$ mediated diagram.
Hence, the overall effect of the VLLs in the muon $g-2$ is inadequate to enhance allowed parameter space
significantly compare to the vanilla type-X 2HDM. 
On the other hand, when the heavy CP even Higgs is heavier than the vector-like lepton, the 
cancellation among $A$ and $H$ mediated process is relatively small. In the next section we will quantify these statements.

Like the electron case, here also the possible charged Higgs diagrams are omitted as
they give very small contribution($\le1-2\%$). The expression for \damu originating from a charged Higgs diagram
is given in ~\cite{Ilisie:2015tra} and the relevant form factor for a $H^\pm$ decay to $W^\pm~\gamma$ via a
vector-like fermion loop is given in~\cite{Song:2019aav}.

\section{Constraints from Precision Observables}\label{sec:precision-obs}
\subsection{Constraints from the $Z$ pole measurements}\label{subsec:z-pole}
We have discussed in the previous section that the mixing of the new leptons with the electron is important to
explain \dae. However, the mixing modifies the coupling of the electron to the gauge bosons, and the precision
measurement at the $Z$ pole~\cite{ALEPH:2005ab} can constrain the mixing. There are three dimension-6 effective
operators which can directly modify the lepton gauge coupling,
\be
\mathcal{L}_{eff} =\frac{1}{\Lambda^2} \left( C_{\phi\ell}^{1,ij} \; \mathcal{O}_{\phi\ell}^{1,ij} \;+\;
                                            C_{\phi\ell}^{3,ij} \; \mathcal{O}_{\phi\ell}^{3,ij}  \;+\;
                                            C_{\phi e}^{ij} \; \mathcal{O}_{\phi e}^{ij}
\right)
\ee
where,
\bea
\mathcal{O}_{\phi\ell}^{1,ij} &=& i\;(\phi^\dagger \overleftrightarrow{D_\mu}\phi)\;(\bar\ell_L^i\gamma^\mu \ell_L^j),\hspace{1cm}
\mathcal{O}_{\phi\ell}^{3,ij} =  i\;(\phi^\dagger {\overleftrightarrow{D_\mu}}^a\phi)\;(\bar\ell_L^i\gamma^\mu \tau^a \ell_L^j),\nonumber\\
\mathcal{O}_{\phi e}^{ij} &=&  i\;(\phi^\dagger \overleftrightarrow{D_\mu}\phi)\;(\bar e_R^i \gamma^\mu e_R^j)
\eea
For a model with a doublet VLL (which couples to the SM singlet) and a singlet VLL (couples to the SM lepton doublet)
we have the following Wilson coefficients,
\be
\frac{C_{\phi\ell}^{1,ij}}{\Lambda^2} =\frac{C_{\phi\ell}^{3,ij}}{\Lambda^2} = -\;\frac{\lambda_E^2}{4 M_E^2}
{\textrm{~~~~and~~~~}}
\frac{C_{\phi e}^{ij}}{\Lambda^2} = +\frac{\lambda_L^2}{2 M_L^2}
\ee
In our scenario the leptons as well as the VLLs couple exclusively to $\Phi_1$. Consequently, the global electroweak
fit for the vector like leptons gives the following limit~\cite{Kannike:2011ng,Crivellin:2020ebi}:
\be\label{eq:limit-lam-L-E}
\dfrac{v_1 \; |\lambda_E|}{M_E}  \leq 0.04 {\textrm{~~~~and~~~~}} \dfrac{v_1 \; |\lambda_L|}{M_L} \leq 0.02 .
\ee
We will satisfy this limit throughout.

\subsection{Constraints from oblique corrections}\label{subsec:oblique}
So far we have seen that the Yukawa couplings which induce mixing of the SM leptons with the VLLs can be constrained
by the precision observables. However, the $Z$ pole observables can not constrain the coupling $\lambda$ and
$\bar\lambda$ which mixes the VLLs among themselves. Interestingly, the mass eigenstates of the heavy charged
leptons depend on these couplings, and they can induce a mass gap between the neutral and the charged component
of the doublet. This can give correction to oblique $T$ parameter~\cite{Peskin:1990zt,Peskin:1991sw}
and can be constrained. Contribution to the $T$ parameter from the VLL is given by~\cite{Lavoura:1992np},
\bea
\Delta T =&& \frac{1}{16\pi s_W^2 c_W^2}  \bigg[\left(c_L^2+c_R^2\right)\theta_+(y_1,y_L) + \left(s_L^2+s_R^2\right)\theta_+(y_2,y_L)
+ 2 c_L c_R \; \theta_{-}(y_1,y_L) \nonumber \\
&&+ 2 s_L s_R \; \theta_{-}(y_2,y_L)) - (s_L^2 s_R^2 + c_L^2 c_R^2) \theta_+(y_1,y_2) - 2 s_L s_R c_L c_R\; \theta_-(y_1,y_2)\bigg].
\eea
Where $y_i=M_i^2/M_Z^2$ and for the heavy neutral particle we denote its mass as $M_L$ disregarding the small
radiative correction. The mixing angle $c_{L/R}\equiv\cos\theta_{L/R}$ etc., are components of the unitary
matrices shown in Eq.~\ref{eq:UL} and \ref{eq:UR}.
The functions $\theta_{\pm}$ are,
\bea
\theta_+(y_1,y_2) &=& y_1 + y_2 - \dfrac{2 y_1~y_2}{y_1-y_2}{~\rm ln}\dfrac{y_1}{y_2} \nonumber \\
\theta_-(y_1,y_2) &=& 2~\sqrt{y_1~y_2}\left(\dfrac{y_1+y_2}{y_1-y_2}{~\rm ln}\dfrac{y_1}{y_2}-2 \right).
\eea

We have scanned the parameters in our model as shown in Tab.~\ref{tab:parameters} and found 
that all points satisfy the constraints.
In our scenario, the mass difference originates from the $vev$ of $\Phi_1$, which is in general small for large
$\tb$. Hence, the constraints from the $T$ parameter can be easily satisfied.

Apart from the vector-like leptons, the scalar sector in 2HDM can also contribute to $T$ parameter. However,
it has been shown that~\cite{Gerard:2007kn,Broggio:2014mna} the oblique corrections from the scalar sector of
2HDM can be minimized by making the charged Higgs degenerate with the heavy scalar or the pseudo-scalar.
We will use this mass spectrum in our analysis.

\section{Results and discussion}\label{sec:result}
\begin{table}[t]
{\footnotesize
\begin{tabular}{|c|c|c|c|c|c|c|}
\hline
Parameters & $v_1\dfrac{|\lambda_{L/E}|}{M_{L/E}}$ & $\lambda, \bar\lambda$       & $M_L$(GeV) &$\Delta M=\dfrac{M_E-M_L}{M_E+M_L}$& $M_A$(GeV)  & $\tb$  \\ \hline
Range      & $(10^{-1}~,~10^{-5})$        & $(-\sqrt{4\pi}~,~\sqrt{4\pi})$ & $(500~,~1000)$&$(0.01~,~0.10)$   & $(30~,~150)$ & $(30~,~100)$ \\ \hline
\end{tabular}
}
\caption{Range of the scanned parameters}
\label{tab:parameters}
\end{table}

In this section, we will present the numerical results for \dae and \damu, which satisfy the precision constraints.
We have scanned the available parameter space shown in Tab.~\ref{tab:parameters}. The Yukawa couplings $\lambda$
and  $\bar\lambda$ are relatively unconstrained due to small $vev$ of $\Phi_1$ as discussed in Sec.~\ref{subsec:oblique}.
Hence we have scanned the full range allowed by perturbativity. The BZ diagram contribution to \damu coming from VLL
loop also depends on the singlet-doublet mixing in the VLL sector as only diagonal couplings from $A(H)\chi_i\chi_j$
appears in the $H$ or $A$ mediated loops. If the singlet and doublet mass parameter, i.e. $M_E$ and $M_L$ are well
separated then the mixing will be small and $A(H)\chi_i\chi_j$ coupling will be mostly off-diagonal and contribution
from VLL loop for \damu will vanish. Hence, for our analysis, we have assumed that the mass difference between the
vector-like masses is small and varied the parameter $\Delta M(=\dfrac{M_E-M_L}{M_E+M_L})$ in a small range which
allows large mixing. For the same reason, the dominant contribution to \damu comes when $\lambda$ or $\bar\lambda$
is relatively large.

\subsection{Results for electron $(g-2)$}

\begin{figure}[t]
\centering
  \includegraphics[width=10cm]{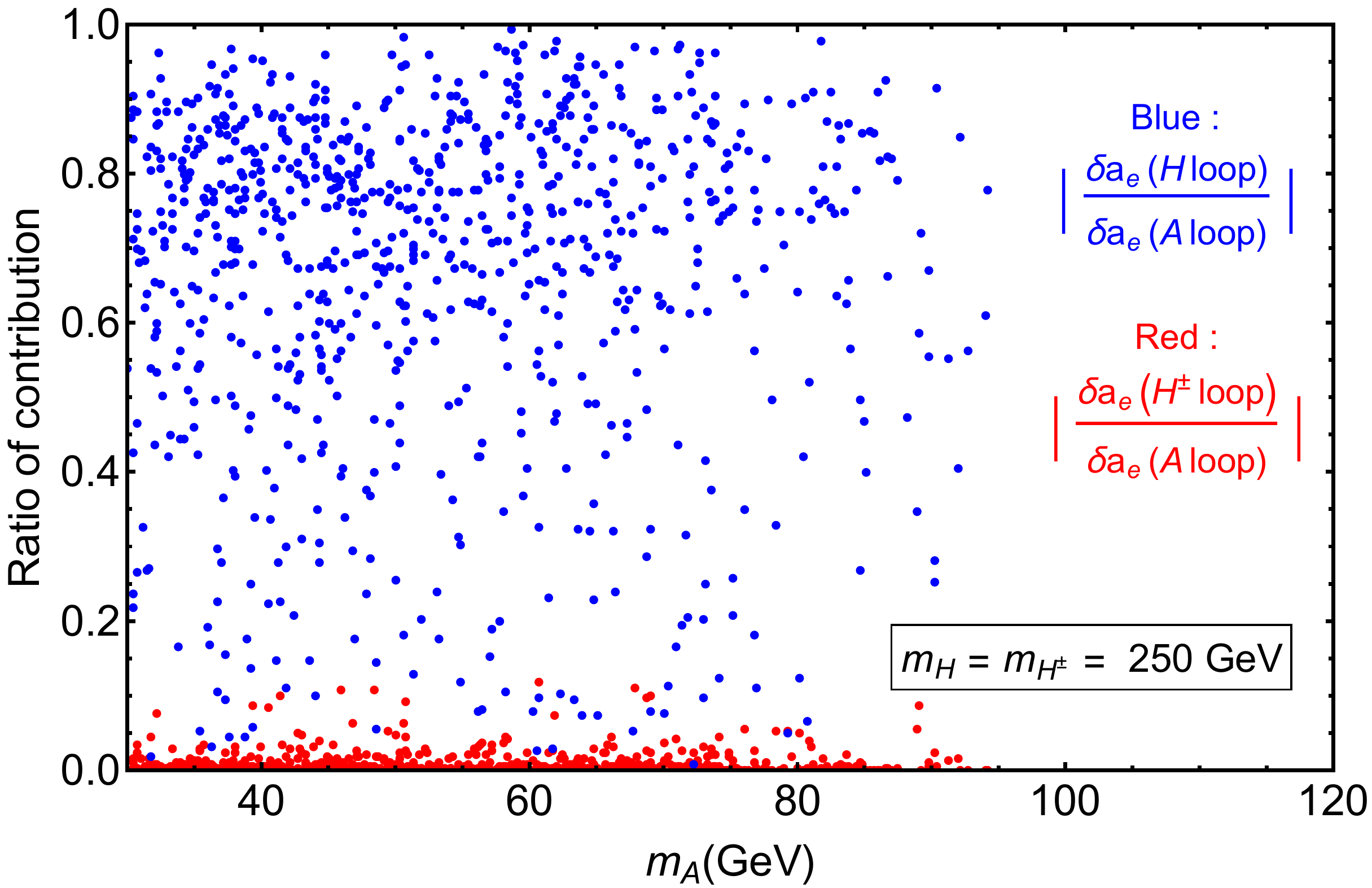}
  \caption{Electron $(g-2)$ contribution coming from the heavy neutral and charged Higgs mediated diagram relative
  to the pseudo-scalar mediated diagram is shown here. We have plotted the absolute value and the contribution coming
  from $H(H^\pm)$ mediated diagram is opposite to(same as) the $A$ mediated diagram.}
 \label{fig:ratio_A_H_del_ae}
\end{figure}

The expressions of new physics contributions for \dae is given in Sec.~\ref{subsec:el-g-2}. The dominant contribution
comes from the pseudo-scalar and heavy Higgs mediated diagram due to $\tb$ enhancement.
The contributions from the gauge boson mediated diagrams are at the percent level or below.

Since the $H$ and $A$ contributions have opposite sign, they will cancel partially.
In Fig.~\ref{fig:ratio_A_H_del_ae} we have plotted absolute value of contribution coming from the $H$ and $H^\pm$
mediated diagram relative to the $A$ mediated diagram as a function of the $A$ mass. Here we set the heavy Higgs
and the charged Higgs mass at 250 GeV for illustration. The contribution from $H$ is not suppressed much and is
a bit smaller than the $A$ contribution. The suppression will be substantial for $m_H \gg M_{L,E}$.
On the other hand, the charge Higgs contribution remains low having no chiral enhancement.

\begin{figure}[t]
\centering
  \includegraphics[width=10cm]{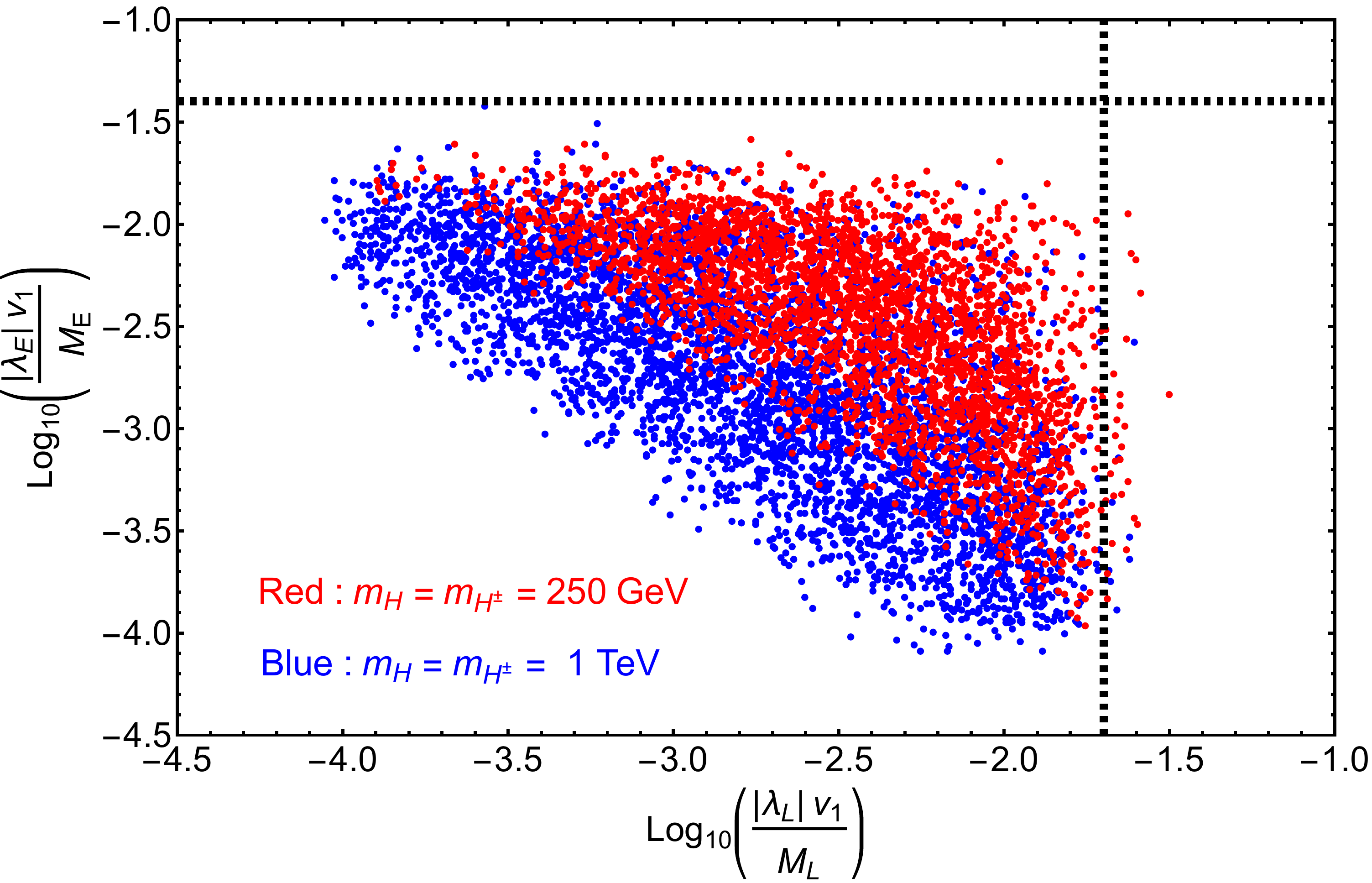}
  \caption{The parameter space in the $\lambda_L - \lambda_E$ plane which can explain the \dae anomaly at 2$\sigma$.
  All the other parameters are varied as shown in Tab.~\ref{tab:parameters}. The black dashed lines are constraints
  coming from $Z$ pole observables.}
 \label{fig:xl_xe}
\end{figure}

In Fig.~\ref{fig:xl_xe} we have displayed the parameter space in $\lambda_L-\lambda_E$ plane which can explain the
electron $(g-2)$ anomaly. The red and blue points illustrate the allowed space for the heavy scalar mass 250 GeV and
1 TeV, respectively. For higher mass, the contribution from $H$ diagram decreases and relatively small values of
$\lambda_L(\lambda_E)$ can explain \dae\!. 
The contribution to \dae dominantly comes from the helicity flipping terms of the
vector-like lepton mass and the contribution is proportional to the factor $\lambda_L \lambda_E \bar\lambda v_1^2/(M_L M_E)$.
This explains the lower bound and correlation among $\lambda_L$ and $\lambda_E$.
The black dashed lines show constraints coming from the $Z$ pole
observations as discussed in Sec.~\ref{subsec:z-pole}. Most of the parameter space remains unconstrained since
the limit coming from $Z$ pole observation is weak.
Apart from precision measurements, perturbativity of the couplings $\lambda_{L/E}(\leq\sqrt{4\pi})$ sets an upper 
    limit on $\dfrac{\lambda_{L/E}~v_1}{M_{L/E}}$. The maximum possible value is $-1.36$ for $\lambda_{L/E}=\sqrt{4\pi},
    \tb=40$ and $M_{L/E}=500$ GeV. Hence, there are no points above the horizontal dotted line which is at $-1.40$ from 
    Eq.~\ref{eq:limit-lam-L-E}. Also, from Fig.~\ref{fig:ma_tb_muon_g-2},
    it is evident that very few points are allowed for small $\tb$ which satisfy \damu. Consequently, there are very few points 
    close to (beyond) the horizontal (vertical) line in Fig.~\ref{fig:xl_xe}. As $\tan\beta$ increases, $v_1$ 
    decreases, which eventually pushes the points towards lower values. Similarly, higher values of $M_{L/E}$ 
will drive the points towards lower value. 

We would like to mention that the helicity flipping terms of the vector-like lepton mass and mixing of 
electron with the vector-like leptons
give rise to new physics contribution in electron mass
$m_e^{NP}\sim \lambda_L \lambda_E \bar\lambda v_1^2/(M_L M_E)$. This contribution comes out to be less than 10\% and do not
alter electron Yukawa coupling or the $h\to ee$ rates which appears in SM+VLL models~\cite{Crivellin:2018qmi}.

\subsection{Results for muon $( g-2)$}
\begin{figure}[t]
\centering
    \includegraphics[width=10cm]{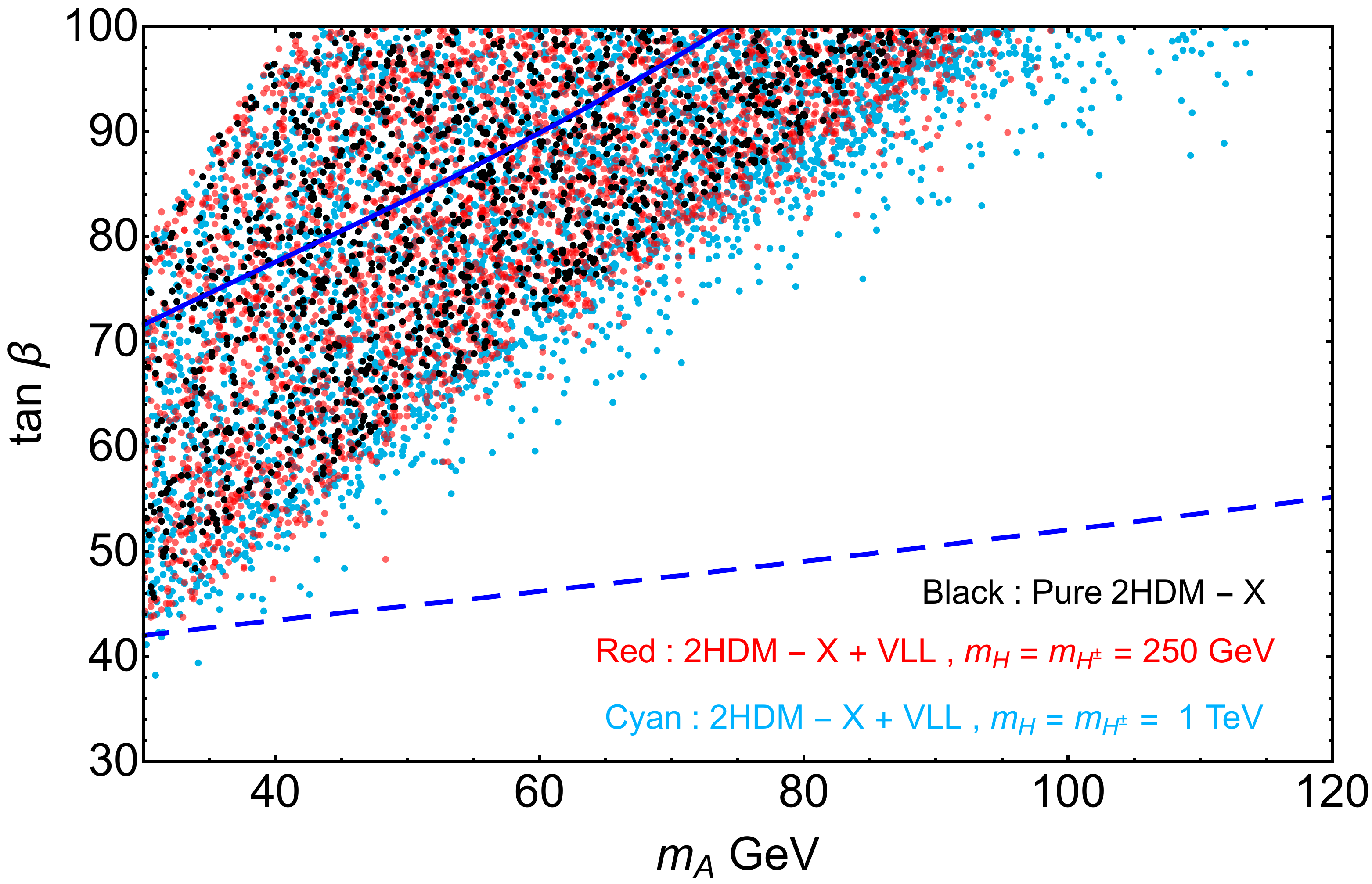}
  \caption{The parameter space in the $m_A - \tb$ plane which can explain muon anomalous magnetic moment at $2~\sigma$.
  The blue solid and dashed lines depict restriction coming from the measurement of $Z\to\ell\ell$ decay.}
  \label{fig:ma_tb_muon_g-2}
\end{figure}

The muon $(g-2)$ anomaly can be explained in the type-X 2HDM model with a light pseudo-scalar when $\tb$ is large.
The dominant contribution comes from the two-loop Barr-Zee diagram with tau loop. In the present model, in
addition to the pure 2HDM contribution, there will be contribution coming from BZ diagram with vector-like leptons
in the loop. In Fig.~\ref{fig:ma_tb_muon_g-2} we showed the parameter space in $m_A -\tb$ plane
which can explain muon anomaly at $2~\sigma$.
 The black colored points in Fig.~\ref{fig:ma_tb_muon_g-2} show the parameter space where muon anomaly can be explained
 in pure type-X 2HDM. In our model additional positive contribution for \damu comes from the pseudo-scalar-VLL loop,
 whereas the heavy scalar-VLL loop contributes negatively. Hence, the parameter space depends on both pseudo-scalar
 mass and heavy scalar mass. For a very heavy $H$(1 TeV) the negative contribution is moderate and larger parameter space
 can explain \damu as shown by cyan colored points. On the other hand, when $H$ is relatively light the cancellation
 among the $A$ and $H$ mediated diagram is large resulting a marginal improvement over the pure type-X parameter space.
 The allowed parameter space for $m_{H/H^\pm} = 250$ GeV is shown in red points.

To illustrate the effect of a heavy scalar in both electron and muon $(g-2)$, we have chosen also $m_{H/H^\pm} 
= 1$ TeV  in  Fig.~\ref{fig:ma_tb_muon_g-2}, which is however tightly constrained by the observation of lepton universality 
in $Z$ and tau lepton decays \cite{Chun:2016hzs,Cherchiglia:2017uwv} .We have shown the $2\sigma$ limits  in blue solid ($m_{H/H^\pm} = 250$ GeV) 
and dashed ($m_{H/H^\pm} = 1$ TeV) curves coming from the observation of $Z\to\ell\ell$ decays.The limit becomes stronger for 
larger hierarchy between the heavy scalars ($H/H^\pm$) and the light pseudoscalar and excludes most of the allowed parameter 
space for $m_{H/H^\pm}=1$ TeV. Here we have not shown the limits from $\tau$ decays as they are much weaker.

\begin{figure}[t]
\centering
    \includegraphics[width=10cm]{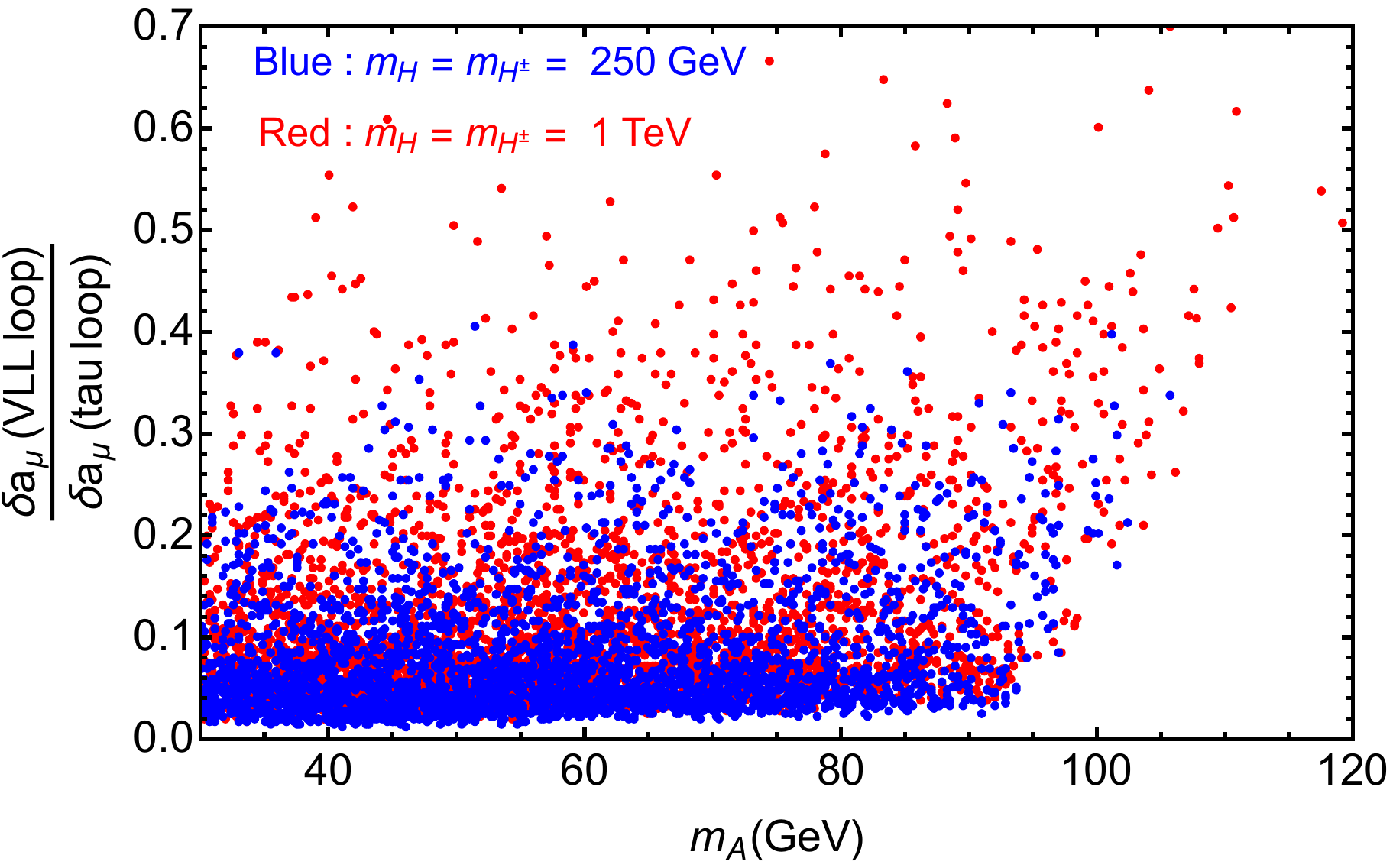}
  \caption{Ratio of the Barr-Zee diagram contribution with VLL and $\tau$ in the loop towards muon $(g-2)$ as a function of pseudo-scalar mass.}
  \label{fig:ratio-tau-vll}
\end{figure}

The contribution coming from the VLL does not alter the parameter space significantly. The ratio of the contributions
coming from the VLL loop and tau loop is shown in Fig.~\ref{fig:ratio-tau-vll} as a function of the pseudo-scalar mass.
The red and blue colored points in Fig.~\ref{fig:ratio-tau-vll} represent the ratio for two different values of
heavy scalar mass. For heavier scalar $H$ the contribution due to vector-like lepton is larger compared to relatively
lighter $H$. When the pseudo-scalar mass is relatively small the tau loop is enough to explain the muon anomaly.
However, for $m_A \sim 100$ GeV the VLL loop need to contribute substantially. The VLL loop contribution remain
small compared to tau loop since the mass insertion in the fermion loop in BZ diagram gives the contribution
proportional to the ratio of chiral mass and vector-like mass $\left(\dfrac{\lambda/\bar\lambda~ v_1}{M_{L/E}}\right)$.

\section{Collider phenomenology}\label{sec:collider}

In the scenario where only a VLL is added to the SM, the charged component of a doublet VLL decays equally to a
lepton and the $Z$/Higgs boson, whereas the neutral component decays to $\ell W$. The branching fraction of a heavy
singlet charged VLL decay into  $W\ell$, $Z\ell$ and $h\ell$ channel is 2:1:1. The search for a vector-like $\tau'$
doublet at the LHC excludes $\tau'$ up to 790 GeV~\cite{Sirunyan:2019ofn} using 77$fb^{-1}$ data. The strongest limit
comes from the di-lepton+tau-jet signal which alone excludes $\tau'$ up to 740 GeV. However, these limits do not directly
apply to our model as the decay channels are completely different as we will discuss now.

In our model, the vector-like leptons couple to the leptophilic doublet $\Phi_1$ and consequently the VLL decays to the
new scalars $A,H$ and $H^\pm$. The coupling of the vector-like leptons to the gauge bosons and light leptons
is $\tb$ suppressed and is negligible. Since the doublet vector-like leptons$\left(\equiv(L^0,L^-)^T\right)$ can be produced through the gauge interaction,
the dominant production channel will via the $W$ boson and we will get the following decay chain:
\be
p\;p\to W^{\ast+} \to L^0 \;L^+ \to (H^+ e^-) (H/A~e^+) \to e^+e^- ~ H^+ ~ H/A.
\ee
The scalars $H^\pm(H)$ decays to $W^\pm A(ZA)$ and $\tau\nu(\tau\tau)$ depending on $\tb$ and mass of the
pseudo-scalar~\cite{Chun:2015hsa,Chun:2017yob,Chun:2018vsn}. The light pseudo-scalar($A$) decays to a pair of taus
since coupling to other leptons is Yukawa suppressed. Hence, depending on the decay channel of the scalars, a plethora
of $\tau$ rich signals along with a pair of high $p_T$ electrons is possible in this model.  Apart from the above
production channel the charged component of the doublet, as well as the singlet, charged vector-like lepton can be
pair produced via the $Z$ boson. However, the production cross-section will be much smaller ($\sim10\%$) than
the $W$ boson channel.

Since the VLL is heavy and decays to a light $A$, our model predicts a very unique signature at the LHC. 
In the decay $L^+ \to e^+\,A$ the transverse momentum for $A$ goes as $p_T(A)\sim \dfrac{m_{L^+}^2-m_A^2}{2~m_{L^+}}$ 
and  the decay products of $A$ will lie within the cone $2m_A/p_T(A)$. Hence, when the vector-like lepton is very heavy 
the light $A$ will be highly boosted, and the tau jets will appear as a single merged jet as the separation will be 
smaller than $\Delta R=0.5$ which is required for tau-jet isolation. Signal of this kind of merged tau pair can be searched by
looking for a lepton in the close proximity of a small radius tau jet. Also the `di-$\tau$ tagger'~\cite{Aad:2020ldt} used 
by the ATLAS collaboration for boosted Higgs searches 
can be useful to look for a leptophilic extended Higgs sector. Hence a dedicated collider study is necessary to look
for the VLL within 2HDM scenario and is beyond the scope of this paper. It is also remarkable that such a light
pseudo-scalar can be readily probed by future linear colliders through the Yukawa process \cite{Chun:2019sjo}.

\section{Conclusion}\label{sec:conclusion}
In conclusion, the type-X 2HDM extended with vector-like lepton doublet and singlet is suggested to explain both
the electron and the muon $(g-2)$ anomaly.  For this, the presence of a light pseudo-scalar is crucial to give a
sizable positive contribution to the muon $(g-2)$ at two-loop and an appropriate negative contribution to the
electron $(g-2)$ with the VLL at one-loop. These features of the model can be tested at the LHC by looking for
a merged $\tau^+\tau^-$ pair accompanied by a pair of $e^+ e^-$. The constraints coming from the precision
observables can be easily satisfied since the new VLL particles couples to the doublet $\Phi_1$ which gets
a small $vev$.

\section{Acknowledgements}
EJC and TM are supported by KIAS Individual Grants PG012504(EJC) and PG073501(TM) at Korea Institute for Advanced Study. 

\providecommand{\href}[2]{#2}
\addcontentsline{toc}{section}{References}
\bibliographystyle{JHEP}

\providecommand{\href}[2]{#2}\begingroup\raggedright\endgroup

\end{document}